# The Chemi-Ionization Processes in Slow Collisions of Rydberg Atoms  with Ground State Atoms: Mechanism and Applications.

A.A.Mihajlov[1], V.A.Sreckovic[1], Lj.M. Ignjatovic[1] and A.N. Klyucharev[2]

*1. University of Belgrade, Institute of Physics, P.O. Box 57, 11001 Belgrade, Serbia*
*2. Department of Physics, Saint-Petersburg University, Ulianovskaya 1, 198504 St. Petersburg, Petrodvorets, Russia*
mihajlov@ipb.ac.rs

In this paper the history and the current state of research of the chemi-ionization processes in atom-Rydberg atom collisions is presented.  The principal assumptions of the model of such processes based on the dipole resonance mechanism, as well as the problems of stochastic ionization in atom-Rydberg atom collisions, are exposed. The properties of the collision kinetics in atom beams of various types used in contemporary experimentations are briefly described. Results of the calculation of the chemi-ionization rate coefficients are given and discussed for the range of the principal quantum number values $5 \leq n \leq 25$. The role of the chemi-ionization processes in astrophysical and laboratory low-temperature plasmas, and the contemporary methods of their investigation are described. Also the directions of further research of chemi-ionization processes are discussed in this paper.

# Introduction: Chemi-Ionization Processes in Thermal Atom-Rydberg Atom Collisions

Atoms in their highly excited states began to attract the researchers' attention as soon as their existence was proven due to the pioneering works of I. Rydberg on atomic spectroscopy which appeared in the literature near the end of the nineteenth century. After the name of their discoverer, the highly excited atomic states were dubbed "Rydberg states", and afterwards the atoms in such states themselves began to be called "Rydberg atoms". Strictly speaking, only truly highly excited states, insignificantly differing from the hydrogen states with the same principal and orbital quantum numbers ($n, l$), should be counted among the Rydberg states. However, in practice an atom $A^*(n, l)$ is being treated as a Rydberg atom if $(n - n_0) \geq 4$, where $n_0$ is the principal quantum number of the outer shell of the atom A in its ground state, and in the case of an atom $He^*(n)$ - for any $n \geq 3$. One should emphasize the fact that at present time even in the laboratory experiments Rydberg atoms with n close to $10^2$ are being explored, whilst the astrophysicists observe the radiation of atoms from the states with n close to $10^3$. With a change of n the parameters characterizing Rydberg states may change by orders of magnitude. Thus with an increase of $n$ of 10 - $10^2$ the lifetime of a Rydberg atom increases $10^5$ times, while at the same time the Rydberg electron's velocity on the corresponding Kepler orbit decreases 10 times etc.

For a long time Rydberg atoms were interesting for certain then actual spectroscopic problems, as well as for the fact that they represent objects which may be explored not only by strictly quantum-mechanical methods, but also quasi-classical, and even purely classical ones. At the present time the interest in processes involving Rydberg atoms is caused by their significance for the fundamental problems of modern atomic physics manifested through the emergence of some new directions of theoretical and experimental research. Here we keep in mind the quantum interferometry [1,2], laser cooling of atoms, in order to create the Bose-Einstein condensate [3-5], and research of the so-called Rydberg-atom matter (RM), i.e. clusters of Rydberg atoms [6-8], from different aspects. Among of them we will mention the investigation of sorption capacity of RM [9] and the creation of some of the forms of RM in the nonideal ultra cold plasmas (see e.g. [10]) .

However, here we want to emphases an other significant reason for the Rydberg atoms studding, namely:  their participation in numerous elastic and particularly non-elastic atomic collision processes which are significant for the weakly ionized gaseous, including here the low temperature plasmas. For such mediums the ionization processes in atom-Rydberg atom collision can be of particularly importance because of their direct influence to the considered gaseous medium ionization degree. The many papers are devoted to the mentioned ionization processes (see for an example review articles [11-16]) .

Among the multitude of ionization processes in the symmetric  $A^{*}(n,l)+A$  and non-symmetric  $A^{*}(n,l)+B$  atom-Rydberg atom collisions ( $n,l$  – the principal and the orbital quantum numbers respectively,  $n \gg 1$ ) there is a prominent class of processes which are connected with the change of the electronic states of the considered systems during the collisions.  As such processes are somewhat similar to chemical reactions, they have come to be known in the literature as "chemi-ionization processes". Apart of that the mentioned class contains the processes of the associative  and Penning – type ionization in symmetric and non- symmetric atom-Rydberg atom collisions, namely:

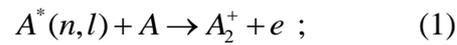

$$A^{*}(n,l)+A \rightarrow A_{2}^{+}+e \; ; \qquad (1)$$

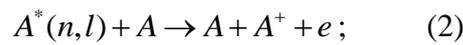

$$A^{*}(n,l)+A \rightarrow A+A^{+}+e \; ; \qquad (2)$$

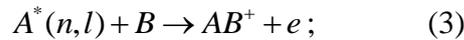

$$A^{*}(n,l)+B \rightarrow AB^{+}+e \; ; \qquad (3)$$

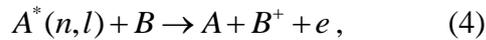

$$A^{*}(n,l)+B \rightarrow A+B^{+}+e \; , \qquad (4)$$

where it is understood that the ionization potential of the atom  $B$  is lower than that of the atom  $A$ .

As one of the ionization channel in atom-Rydberg atom collisions can be treated also the following processes

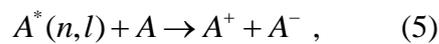

$$A^{*}(n,l)+A \rightarrow A^{+}+A^{-} \; , \qquad (5)$$

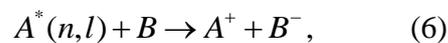

$$A^{*}(n,l)+B \rightarrow A^{+}+B^{-} \; , \qquad (6)$$

in which the positively and negatively charged ions are created, under the condition of the existing of the corresponding stable negative atomic ion. In this paper the history and the current state of research on just the processes of the types (1) - (6) is exposed.

Although some of the chemi-ionization processes in atom-Rydberg atom collisions had already been described in [17], their intensive experimental research began somewhat later, when the development of laser technique made it possible to achieve precise transitions of atoms to the chosen Rydberg states [14, 18-21]. Since from the experimental aspect operation was easiest when the atoms' ionization potentials were small, chemi-ionization processes were studied particulary thoroughly in the cases of alkali metal atoms ( $Li, Na$ , etc.). As it is well known, such processes remain a subject of experimental research even now [22-27]. Large atom ionization potentials were not an insurmountable obstacle however, and the processes with atoms of some rare gases ( $He, Ne$ , etc.) were also experimentally researched during several years (see for example [28-33]).

The content of this paper is distribute in eight sections. The first three of them are devoted to the describing of the dipole resonant mechanism of chemi-ionization processes, and the corresponding methods of the determination of such processes rate coefficients. In other five sections are shown the existing theoretical and experimental results concerning the chemi-ionization processes, their roll in the low temperature laboratory and astrophysical plasmas, as well as the methods of the investigation of such processes. Finally, at the end of this paper the directions of further research are discussed.

## 1. Dipole Resonant Mechanism of Chemi-Ionization:

Since the chemi-ionization processes are treated already for a long time on the bases of the so-called dipole resonant mechanism (DRM), we will describe shortly its history and the main features. This mechanism was introduced in [34] in order to explain the properties of the distribution of highly excited (Rydberg) atomic states in a non-equilibrium helium plasma with $N_e / N_a \approx 10^{-4}$ and $T_e \gg T_a$, which were noticed in [35]. An obstacle emerged from the fact that a Boltzmann distribution of these populations with a temperature $T_e$ was expected, and the distribution found instead looked like a Boltzmann one, but with an effective temperature average between $T_e$ and $T_a$. This suggested the idea that the influence of atom-Rydberg atom collisions on the populations of excited atoms in the considered case was quite comparable to the influence of electron-atom collisions.

The main problem with this fact was the impossibility of explaining it by the well-known Fermi's semi-classical mechanism of non-elastic atom-Rydberg atom collision processes described in [36]. This mechanism implied a dominant role of atom-atom collisions with $\rho \sim r_n$, where $\rho$ is the collision parameter, and $r_n$ is the mean radius of an atom in a Rydberg state with the principal quantum number n. However, in [36] not only collisions with $\rho > r_n$ but also with $\rho < r_n$ were taken into account, in which case a part of the collision trajectory lay inside the outer electron's orbit of the Rydberg atom. Such collisions were taken into account so that an additional factor of influence on the outer electron could be introduced, namely its interaction with the dipole moment induced in the projectile atom by the Rydberg atom core. The fact that the magnitude of the induced dipole moment increases with a decrease of the distance R between the Rydberg atom core and the projectile atom lead to the idea that the influence of the said electron-dipole interaction should be studied within the entire region $\rho < r_n$. As a result, a necessity arose to examine atom-Rydberg atom collisions also at $\rho \ll r_n$, where in one part of the trajectory ($R \ll r_n$) the inner subsystem, that is, the Rydberg atom core plus the projectile atom, must be treated as a quasi-molecular complex. As for the electron component, such a treatment implies that it is to be described through a superposition of the corresponding molecular ion's wave functions, that is $A_2^+$ in the case of a symmetric $A^*(n) + A$ collisional system, and $AB^+$ in the case of a non-symmetric $A^*(n) + B$ system. In both cases those states of the molecular ions are taken into account which are adiabatically correlated to the corresponding states of the considered ion-atom subsystem at $R \to \infty$.

In [34], where non-elastic processes in $H^*(n) + H(1s)$ collisions ($n \gg 1$) were considered, the ion-atom subsystem was described by means of the ground ($1\Sigma_g^+$) and the first excited ($1\Sigma_u^+$) electronic state of the molecular ion $H_2^+$. These two states are denoted here respectively as $|g; R>$ and $|u; R>$, and the potential curves corresponding to these states - as $U_g(R)$ and $U_u(R)$, respectively. It is considered that during the collision $R = R(t)$, where $t$ is time. Accordingly, the state of the ion-atom subsystem $H^+ + H$ may be represented as a superposition:

$$| \Psi(R(t)) >= a_g(t) | g; R(t) > \exp[-\frac{i}{\hbar} \int_{-\infty}^{t} U_g(R(t))dt] + a_u(t) | u; R(t) > \exp[-\frac{i}{\hbar} \int_{-\infty}^{t} U_u(R(t))dt], \quad (7)$$

where it is taken that $| a_g(t) |=| a_u(t) |=1/\sqrt{2}$. Such a way of describing does make sense, as the region R in which the values of the splitting-term $U_{sp}(R)$, defined by the relation

$$U_{sp}(R) = U_u(R) - U_g(R), \quad (8)$$

significantly change, in the region $R << r_n$. This guarantees that the said region $R$, which can be treated as the reaction zone, lies deep inside the outer electron's orbit of the Rydberg atom $H^*(n)$. Therefore in this region the first two (principal) members of decomposition of the potential of the outer electron's interaction with the inner ion-atom subsystem are : Coulomb and the dipole one. This allows for the Rydberg states of the outer electron to be still determined in a Coulomb potential and to be characterized by a principal quantum number $n$ and an orbital number $l$, and for the electron-dipole interaction to be treated as the cause of transition between those states, including the transitions with a change of the principal quantum number, that is, the so-called (n-n')-mixing.

In [34], it was shown that this mechanism provides values of the cross-sections for (n-n')-mixing in the region $n \sim 10$ greater by at least an order of magnitude than the ones obtained by means of Fermi's mechanism [36]. This is due to the fact that in Fermi's mechanism an induced dipole moment $d = e \cdot l$ appears, with $l \ll a_0$, while in the above mechanism the dipole moment

$$D(t) =< g; R(t) | \hat{D}(R(t)) | u; R(t) >, \quad (9)$$

is used, where $\hat{D}(R)$ is the operator of dipole moment of the molecular ion $H_2^+$. It can be shown that in the region $R > 0.5 \cdot a_0$ we can take that

$$D(t) \cong \frac{e \cdot R(t)}{2}, \qquad (10)$$

where e is the modulus of the electron charge. In [34] it was determined that the considered process of (n-n')-mixing has a resonant character. Namely, from Eqs. (7) and (9) the dipole moment $D(t)$ may be represented in the form

$$D(t) = \frac{R(t)}{2} \cdot \cos\left(\int_{-\infty}^{t} \omega(R(t')) \cdot dt'\right), \qquad \omega(R(t)) \equiv \frac{1}{\hbar} \cdot U_{sp}(R(t)), \qquad (11)$$

where $\omega(R)$ is the frequency of oscillations of $H + H^+$ subsystem's dipole moment. Consequently, the transitions $n \rightarrow n' \neq n$ dominantly take place in narrow neighborhoods of the resonant points $R_{n;n'}$, which are the roots of the equation

$$\omega(R) \equiv \frac{1}{\hbar} \cdot U_{sp}(R) = \omega_{nn'} \equiv \frac{1}{\hbar} \cdot |\varepsilon_{n'} - \varepsilon_n|, \qquad (12)$$

where $\omega_{nn'}$ are the frequencies corresponding to the outer electron's transitions between the states with energies $\varepsilon_n$ and $\varepsilon_{n'}$. The facts: that the probability of the outer electron's transition $n \rightarrow n' \neq n$ is determined by $|D|^2 = D^2$, where $D \gg 1$, and that the transitions themselves have a resonant character, account for the great differences in efficiencies between the Fermi's and the just described DRM. Somewhat later, in [37], using the same mechanism as in [34], ionization processes in symmetric $Na^*(n, l) + Na$ collisions were considered. These processes were caused by the outer electron's transitions from a bound state with energy $\varepsilon_n < 0$ to one of free (continuum) states with energy $\varepsilon_k = p^2/2m$, where $p$ is the momentum, and $m$ - the electron mass. It was determined that the considered processes have a resonant character too, as the electron transitions $\varepsilon_n \rightarrow \varepsilon_k$ were also dominantly bound to narrow neighborhoods of the corresponding resonant points $R_{nk}$, determined as the roots of the equation

$$\omega(R) \equiv \frac{1}{\hbar} \cdot U_{sp}(R) = \omega_{nk} \equiv \frac{1}{\hbar} \cdot (\varepsilon_k - \varepsilon_n),$$

and that this mechanism in the case of the considered ionization processes is very efficient too, just as in the case of (n-n')-mixing.

In the same period the investigation of processes of (n-n')-mixing in $A^*(n) + A$ collisions was continued [38], which lead to a modification of the method of description of these processes. Namely, in [34], as well as subsequently in [37], the states of the system $A^*(n) + A$ in the reaction zone, were sought as superpositions of states $|\Psi(R(t))>|n, l>$, where $|\Psi(R(t))>$ is defined by (7), and $|n, l>$ is one of the Rydberg states of the outer electron. However, the results obtained in [38] suggested that the states of the system $A^*(n) + A$ be sought as superpositions of two groups of states: $|g; R>|n', l'>$ and $|u; R>|n'', l''>$, where the principal quantum numbers $n', l', n'', l''$ take all their possible values independently from one another. Mutual ortogonality of the states from the first and the second group is automatically provided by ortogonality of the states $|g; R>$ and $|u; R>$. Already in [39] these results were applied to the chemi-ionization processes in $A^*(n) + A$ and $A^*(n) + B$ collisions. In that paper the chemi-ionization processes were treated as a result of the electronic state of the collisional system decay under the influence of the above described electron-dipole interaction of the outer electron and the inner ion-atom subsystem $A^+ + A$ or $A^+ + B$. The states $|2, R>|n, l>$ and $|1, R>|k, l'>$ were treated respectively as the initial and the final state, where $|1, R>$ and $|2, R>$ are respectively the ground and the first excited electronic state of the ion-atom subsystem, asymptotically correlated to the corresponding states of the same subsystem ($A^+ + A, A + B^+$; $A^+ + A, A^+ + B$) at $R \rightarrow \infty$, and $|k, l'>$ is one of the free (continuum) states of the outer electron with an energy $\varepsilon_k$.

The papers [37,39] practically opened two directions of description of chemi-ionization processes. Each of them implies a semi-classical treatment of the atom-Rydberg atom collisional system, the same electron-dipole interaction as the cause of the change of the outer electron's state, but with different ways of the electron's transition from bound states to the continuum state. A way of describing chemi-ionization processes, similar to that presented in [39], is treated here as the "decay approximation" based on the dipole resonant mechanism. This approximation is considered in more detail in the next section. The other way, similar to that of [37], is treated here as the "stochastic (diffusional) approximation" based on the same resonant mechanism, and is considered in Section 3.

## 2. Chemi-Ionization Processes in the Decay Approximation

The manner of application of the dipole resonant mechanism in the decay approximation according to [39] is characterized by several basic elements. As first it is assumed that the relative motion of the nuclei in $A^*(n) + A$ or $A^*(n) + B$ collisions is described by a classical trajectory characterized by the collision parameter $\rho$ and the initial (collisional) energy E, under condition that

$$\rho \ll r_n \sim n^2, \qquad (13)$$

where $r_n$ is the mean radius of the atom $A^*(n)$, as it is illustrated by Fig. 1, and that the trajectories are considered which, for given $n, \rho$ and $E$, pass through the chemi-ionization zone, that is, the region

$$R \le R_{i;n}, \qquad (14)$$

where $R$ is the internuclear distance, and the boundary value $R_{i;n}$ is determined by the ionization energy $I_n$ of the atom $A^*(n)$ and the corresponding parameters of the ion-atom subsystem ( $A^+ + A$ or $A^+ + B$ ).

Within this approximation the ion-atom subsystem is described by means of the electronic states of the corresponding molecular ions, $A_2^+$ or $AB^+$ , denoted here as $|1, R>$ and $|2, R>$, supposing that $I_B < I_A$, where $I_A$ and $I_B$ are the ionization potentials of the atoms $A$ and $B$ , and that

$$U_{12}(R) \equiv U_2(R) - U_1(R) > 0, \qquad (15)$$

where $U_1(R)$ and $U_2(R)$ are the potential curves corresponding to the respective states $|1, R>$ and $|2, R>$. Here $|1, R>$ is the ground electronic state of the molecular ion, or, in the general case, one from a group of states asymptotically correlated to the same state of the subsystem $A^+ + A$ or $A + B^+$ at $R \to \infty$, to which the ground state is correlated.

Similarly to that, here $|2, R>$ is the first excited electronic state of the molecular ion, or, in the general case, one from a group of excited states asymptotically correlated to the same state of the subsystem $A^+ + A$ or $A + B^+$ at $R \to \infty$, to which the first excited state is correlated.

In the decay approximation the chemi-ionization processes are treated as a result of the transition

$$|2, R> |n, l> \to |1, R> |k, l'>, \quad (16)$$

of the complete system from an initial state $|2, R> |n, l>$ to a final state $|1, R> |k, l'>$, which is caused by DRM. This event can be treated as a simultaneous transitions of the outer electron and of the ion-atom subsystem, namely: $|n, l> \to |k, l'>$ and $|2, R> \to |1, R>$ respectively, which is also illustrated in Fig.1.

It is assumed that transitions (16) predominantly take place within narrow neighborhoods of the resonant points $R_{nk}$ which are the roots of the equation

$$\varepsilon_{nk} \equiv \varepsilon_k - \varepsilon_n = U_{12}(R), \quad (17)$$

where $\varepsilon_k = (\hbar k)^2 / 2m$ and $\varepsilon_n = -I_H / n^2$ are the energies of the outer electron in the initial (bound) and the final (free) state, and the splitting term $U_{12}(R)$ is defined by (15). During the collision the chemi-ionization process is described by the ionization rate $W_{i;n}(R)$, which implies that, for a given $R$ the outer electron passes to the free state with energy $\varepsilon_k(R) = U_{12}(R) + \varepsilon_n$. In accordance with [39] the quantity $W_{i;n}(R)$ can be presented in the form

$$W_{i;n}(R) = \frac{4}{3\sqrt{3}n^5} \cdot D_{12}^2(R) \cdot G_{nk}, \qquad D_{12}\left(R\right) = <2;R \,|\, \hat{D}(R) \,|\, 1;R>, \qquad (18)$$

where $\hat{D}(R)$ is the operator of the dipole momentum of the considered ion-atom subsystem, and $G_{nk}$ is the corresponding (bound-free) Gaunt factor (see [40]).

It is understood that the chemi-ionization process takes place within the interval $(-t_0, t_0)$, where $-t_0$ is the moment of the collision system's entrance into the reaction zone defined by the relation (14), where $R_{i;n}$ is a root of the equation $U_{12}(R) = |\varepsilon(R)|$, $t_0$ - the moment of leaving the zone, and $t = 0$ is taken to be the moment when the collision system reaches the minimal internuclear distance $R_{\min} \equiv R_{\min}(\rho, E)$. In accordance with [39] the probability $P_{i;n}(t; \rho, E)$ of the chemi-ionization process in an $A^*(n) + A$ or $A^*(n) + B$ collision with given $\rho$ and $E$ is expressed by $P_{i;n}(t; \rho, E) = p_0(2) \cdot p_{i;n}(t; \rho, E)$, where $t$ is the time, $p_0(2)$ - the probability of the collision system's entering the reaction zone just in the initial state $|2, R> |n, l>$, i.e. $p_0(2) = 1/2$ in the symmetric $p_0(2) = 1$ in non-symmetric case, and $p_{i;n}(t; \rho, E)$ - the probability of this state's decay, which must satisfy the equation

$$\frac{dp_{i;n}(t; \rho, E)}{dt} = \left[1 - p_{i;n}(t; \rho, E)\right] \cdot W_{i;n}(R(t)), \qquad -t_0 \le t \le t_0, \qquad (19)$$

with an initial condition

$$p_{i;n}(-t_0; \rho, E) = 0. \qquad (20)$$

One can see that the solution of this equation can be presented in the form

$$p_{i;n}(t; \rho, E) = 1 - \exp(I_{i;n}(t; t_0)), \qquad I_{i;n}(t; t_0) = \int_{-t_0}^{t} W_{i;n}(R(t))dt. \qquad (21)$$

From here it follows that the probability of chemi-ionization for given $\rho$ and $E$, e.g. $P_{i;n}(\rho, E) = p_{in} \cdot p_{i;n}(t = t_0; \rho, E)$ is defined by relations

$$P_{i;n}(\rho, E) = p_{in}[1 - \exp(2I_{i;n}(t_0))], \qquad I_{i;n}(t_0) = \int_{0}^{t_0} W_{i;n}(R(t))dt. \quad (22)$$

Than, taking that $dt = dR / v_{rad}(R; \rho, E)$, where $v_{rad}(R; \rho, E)$ is the internuclear radial velocity, we can presented $P_{i;n}(\rho, E)$ in the form

$$P_{i;n}(\rho, E) = p_{in}[1 - \exp(2 \cdot \int_{R\min}^{R_{i;n}} \frac{W_{i;n}(R)dR}{v_{rad}(R; \rho, E)})], \qquad v_{rad}(R; \rho, E) = \sqrt{\frac{2}{\mu}[E - U_2(R) - \frac{E\rho^2}{R^2}]}, \quad (23)$$

where $W_{i;n}(R)$ is given by Eq. (18).

After that, the cross-section $\sigma_{ci;n}(E)$ for the considered chemi-ionization process and the corresponding rate coefficient $K_{ci;n}(T)$ can be presented by means the usual expressions

$$\sigma_{ci;n}(E) = \int_{0}^{\rho_{\max}(E,n)} P_{i;n}(\rho, E) \cdot 2\pi\rho d\rho, \quad (24)$$

$$K_{ci;n}(T) = \int_{E_{\min}(n)}^{\infty} \sigma_{ci;n}(E) \cdot f(E;T)dE, \quad (25)$$

where $P_{i;n}(\rho, E)$ is given by Eqs. (18) and (23), the parameters $\rho_{\max}(E,n)$ and $E_{\min}(n)$ are determined by the behavior of the potential curve $U_2(R)$ and the splitting term, and $f(E,T)$ is the corresponding distribution function over $E$ at a given temperature $T$, which depends on the considered conditions (a single beam, crossed beams, a cell).

From the expression presented it follows that sense of $\sigma_{ci;n}(E)$ and $K_{ci;n}(T)$ is determined by the one of the Gaunt factor $G_{nk}$ in Eq. (18). Namely, in the case when $G_{nk}$ is obtained by the averaging of the partial Gaunt factors $G_{nk}(l)$ over all possible $l$, from $l = 0$ to $l = n-1$, $\sigma_{ci;n}(E)$ and $K_{ci;n}(T)$ have the sense of the mean characteristics for the whole shell with given $n$. In other case, when $G_{nk} = G_{nk}(l)$ the quantities $\sigma_{ci;n}(E)$ and $K_{ci;n}(T)$ are the cross-section and rate coefficient for the chemi-ionization processes with given $n$ and $l$. Also it is necessary to remember that in the principle the energies of the electron Rydberg states $\varepsilon_n = 1/[n - \Delta n(n,l)]$, where $\Delta n(n,l) > 0$ sometimes can't be neglected.

Let us note that by now only one article [41] was devoted to completely quantum-mechanical describing the chemi-ionization processes in atom-Rydberg atom collisions. In that article such processes in $Cs^*(n) + Cs$ collisions were considered in the quasi-classical approximation at $T \approx 1000K$, on the basis of DRM. The results obtained in [41] are in a good agreement with all the corresponding results obtained in the above described semi-classical approximation (see for example [42]). Unfortunately, the additional approximation used in [41] makes impossible applying the quasi-classical method developed in that article to the very important region of extremely low temperatures.

# 3. Stochastic Ionization in Atom-Rydberg Atom Collision

The results of research in the field of non-linear mechanics show that dynamic-chaos regime should be considered a typical, rather than an exceptional, situation in the dynamics of Hamiltonian systems even with few degrees of freedom. In the quasi-classical variety of the theory of collisions the chaotic regime may come as a consequence of local instability of the optical electrons' trajectories with respect to small perturbations [43]. In the general case there are always regions of the phase space and the interacting particles' parameters for which the dynamics of a Hamiltonian system is stochastic, the transition from the integrable problems of classical dynamics to the systems with chaos is accompanied by appearance of embryo-regions of chaos. With one-dimensional problem and time-dependent influence of the external perturbation development of dynamic chaos is possible [43].

The theory cited above did not take into account the possibility of multiplicity of quasi-intersections of the initial term of Rydberg molecular complex $(A^* + A)$ with the adjacent term prior to its reaching the border of the continual spectrum. At the same time the DRM theory [34] initially considered non-elastic processes of intermixing Rydberg states. However, the attempt to take that possibility into account by the known means, traditional within the physics of atom-atom collisions, do not lead to a foreseeable result due to the multiplicity of such quasi-intersections. The distance between the terms of energy of a Rydberg atom is ($\delta\mathcal{E} \sim n^{-3}$), while the bond energy is $\mathcal{E}_0 \sim (2n^2)^{-1}$. Thus, $\delta\mathcal{E} / \mathcal{E}_0 \sim n^{-1} \ll 1$. In the paper [44] this lead to the idea of a diffusional approach to the problem of atom-Rydberg atom collisions. By this we mean diffusion of Rydberg electron (RE) over the excited quasi-molecul's energy terms during a single collision. In the experiment on exposing a gas target to laser radiation [45] the mechanism of RE diffusion in the electromagnetic field was proposed as a way of explaining an anomalously large photo-ionization cross-section and a critical value of field energy was obtained at which the RE's motion attains stochastic character, see also [46]. Subsequent papers [23], [47] lead to the conclusion that RE's stochastic diffusion has resonant character and happens under the influence of internal non-linear dynamical resonances, taking place when overtones of the frequency of the RE's motion along a Kepler orbit coincide with the frequency of an electronic transition in the excited atom. As a result, the RE's motion over the energy spectrum turns into a random walk over over the net of intersecting terms. This leads to a temporal evolution of population distribution $f(n,t)$ of Rydberg atoms.

We shall point out the main propositions of the stochastic-diffusion model:

I. Characteristic times of stochastization.
1. Collisional dynamics of the system of colliding atoms determines the dependency of the internuclear distance $R(t)$ in the Rydberg complex.
2. Factorization of the Fokker-Planck equation (with respect to $R$ and $n$) means introduction of the "effective time" $t_{eff}$.
3. The overall effective time of the collision $\tau$ with a fixed impact parameter $\rho$ equals the time within which $R(t)$ changes from $R = \infty$ to the turning point $R_{min}$.
4. The notion of mean effective time of diffusion $< t_{eff} >$, within which the RE diffuses from the initial level with $n_0$ to the ionization limit.

II. The region of existence of dynamic chaos.
1. For stochastic instability of the RE's motion to develop, the perturbation must exceed certain critical value which can be estimated through the so-called resonance-overlap criterion. The range of values of $n_{eff}$ within which stochastic diffusion of RE is possible is determined by the criterion from [48].

2. At the lower boundary of the stochastization range of $n_{eff} \geq N_{min}(R)$ diffusion is limited due to the increase of energy gap between the adjacent terms of excited atomic states.

3. The corresponding upper boundary $N_{max}(R) \geq n_{eff}$ is connected to the appearance of effective photo-ionization channel of the Rydberg quasi-molecule.

4. The effective time of diffusion $< t_{eff} >$ must be smaller than the time of the collision itself $\tau$.

Thus, stochastic diffusion of a RE during a single collision should manifest primarily within the region of slow collisions.

**Stochastic diffusion under conditions of Förster resonance.** Large values of dipole moments for transitions between Rydberg states of an excited atom, cased by their high polarizability $\alpha \sim n^7$ lead to appearance of non-linear effects in weak electric fields. One of them is the effect of double Stark resonance (Förster resonance) [49]. This corresponds to the case when an l-series level lies between two levels of $l-1$ series ( for an example some $P$-state between two $S$ - states ), much like the scheme of levels of a three-dimensional quantum oscillator. According to Bohr's second correspondence rule, in such a system it is mostly transitions to adjacent levels that are allowed. By changing the magnitude of displacement of an intermediate $nS$ level by means of electric field, it is possible to pass from the case of two-photon transition between $(n-1)P - nP$ states of the Rydberg atom to its $(n-1)P - nS - nP$ step variety. This leads to an increase of probability of $(n-1)P - nP$ transition and, as a result, to a decrease of the lifetime of the nP state, which leads to a decrease of diffusional flux through the $nP$ level and prolongation of the stochastic ionization time. In figures (3a,b) the results of calculation are presented of time-dependency of RE's energy in stochastic-diffusion regime and blockage of optical transitions - the case of Förster resonance [50].

# 4. Features of Atom-Collision Kinetics under Conditions of Atomic Beams

In the paper [51] attention was drawn to the prospect of using atomic beams for investigation of non-elastic processes within the volume of the beam itself. This was realized for the case of chemi-ionization process at binary collisions of resonance-excited sodium atoms [52]. Afterwards analogous work involving usage of single and crossed atomic beams was done in a line of laboratories, see e.g. [21], [53]. On those occasions it turned out that the difference between the results of beam experiment [21] and the data obtained under conditions of gas cell and of perpendicular crossed beams significantly exceeded possible errors. In [54] this situation was named "the sodium paradox". As far as we know, the first explanation of the cause of discrepancy among the published results was given in [55], where attention was drawn to the significant difference between the respective distribution functions of relative velocity of atoms in the two cases. We shall examine briefly the distribution functions of particles over the velocities of their relative motion $v_{cl}$ in the beams of different types, which will come in useful later on at the analysis of experimental data from various laboratories.

Under conditions of a gas cell (low-temperature plasma) the case is realized of isotropic distribution of the colliding neutral particles over their relative speed. In contemporary facilities divergence of a single effusion beam does not exceed several degrees, so further on we shall consider collimated beams as the first approximation. At optical excitation under conditions of beam experiments the shape of the distribution function of excited atoms over the velocity of their relative motion depends on the angle of the beams' intersection. The graph of the function $f(v)$ is presented in Fig.4. A single atomic beam of effusional type may be considered a special case of two beams intersecting at an angle $\theta = 0^{o}$ (the approximation of catching-up beams). In this situation for endothermic processes cases of anomalously small rate coefficients may be realized [55].

The calculation gives a decrease of the mean collision velocity of the particles in the beam at $\theta \approx 10^{o}$ of 2.7 times with respect to the gas cell, which decreases the mean collision energy approximately for an order of magnitude. Such collisions were dubbed sub-thermal in the literature $< E_{cl} > \sim 10^{-3} eV$ , and may be considered as an intermediate case between thermal $< E_{cl} > \ll (0.1-1)eV$ and cold collisions at $<E_{cl}> \ll 10^{-3} eV$ in the experiments on laser cooling of particles. At comparable temperature of the beam sources the following relations between the relative velocities of particles are known:

$$<v_{cl}>_{sb} \cdot <v_{cl}>_{cell}^{-1} = 0.37;$$
$$<v_{cl}>_{cb} \cdot <v_{cl}>_{cell}^{-1} = 1.04; \qquad (26)$$
$$<v_{cl}>_{gb} \cdot <v_{cl}>_{cell}^{-1} = 0.5.$$

where $<\ldots>$ denotes the result of the averaging under the corresponding conditions. Here we have the following cases:

in a gas cell - $<\ldots>_{cell}$, in a single beam - $<\ldots>_{sb}$, in beams intersecting at an angle of 90° - $<\ldots>_{cb}$ and 180° - $<\ldots>_{cb1}$, in a gasodynamical beam [56], with a mass velocity $v_m \gg v_0 - <\ldots>_{gb}$, $v_0 = (2kT/\mu)^{1/2}$, and $\mu$ is the reduced mass of the particles.

In the case of gasodynamical beam it is supposed that the motion of the atoms is single-directed, i.e. that there are no components of the speed perpendicular to the direction of the beam's motion. The first publications concerning this belong to the year 1984 [55], see also [27].

These relations may be used at the comparative analysis of the experimental data in order to estimate the parameters of associative ionization involving excited atoms, such as the smallest inter-atomic distance at which an auto-ionization decay of the initial state of the excited quasi-molecule is possible, $R_n$, auto-ionization width of the decay covalent term at $R < R_n$ and the energy threshold of the reaction [56].

# 5. Atom-Rydberg Atom Collisions. The Theoretical and the Experiment data

**Symmetric collisions.** We shall present examples of data on chemi-ionization within the range $5 \leq n_{eff} \leq 25$, obtained under conditions of gas cell and of atomic crossed beams. The theoretical results for $Na$ are presented in Fig. 5. In the dependency $K_{ci;n_{eff}}(n_{eff})$ a wide maximum is conspicuous at values $n_{eff} \sim 10$, $T \approx 10^3 K$, and the maximal $K_{ci;n_{eff}}(n_{eff}) \approx 10^{-9} cm^3 s^{-1}$ characteristic of endothermic reactions. The branch coefficient $X^{(a,b)}(n_{eff})$ at values $n_{eff} \approx 5$, reaching about 25% at the region of the maximum, increases to 75% for $n_{eff} = 5$, which is significantly greater than the data for binary collisions of metastable atoms of inert gases known from the literature. In figure 6 from [57] the results are given for $n_{eff}\,^2S$, $n_{eff}\,^2P$, $n_{eff}\,^2D$ of sodium atoms with orbital quantum number values of 0, 1 and 2 respectively. It can be seen that the difference between the theoretical data, obtained in the approximation of stochastic dynamics (single-trajectory model) and the data from beam experiments exceeds the error margin of the experiment. The results on $Na$, calculated in [40] are in accordance with the experiment [58], within the limits of its error for $n^2S$ states of the Rydberg atom. In the stochastic approximation of the theory it may be connected with the right way of inclusion of the correlation between the electronic and the nuclear component of the Rydberg cluster's energy. Another possibility for correction concerns the fact that in the crossed-beam experiment the registered ion signal represents the sum of the signal coming from the collisions of particles from the two beams, and the signal induced by the collisions of particles within each beam. This circumstance was pointed out in [59] (see Fig. 7). As it was noted above, the stochastic approximation of the theory is not applicable to the excited states adjoining the block of Rydberg atoms from below. In that case individual features of the excited clusters' potential curves should be taken into account, as well as the possibility of their quasi-intersection, which opens additional channels of ionization.

Lately, the chemi-ionization processes of the types (1) and (2) where $A$ is the ground-state atom of the one of the rare gases become a subject of theoretical and experimental investigation again [60]. Firstly, the associative ionization processes (1) with $A = Ne$ are being taken into account here. In the case of the chemi-ionization processes $Ne^*(n) + Ne$ the electrons of the outer shell of the ground-state atom are in $p-$states, and consequently, the ion-atom subsystem $Ne^+ + Ne$ can enter the reaction zone not only in one of two $\Sigma-$states, but also in one of four $\Pi-$ states. Therefore the total associative ionization rate coefficients $K_{ci;n}(T)$ for the processes (1) with $A = Ne$ have to be taken in the form

$$K_{ci;n}(T) = \frac{1}{3} K_{\Sigma;n}(T) + \frac{2}{3} K_{\Pi;n}(T),$$

(27)

where $K_{\Sigma;n}(T)$ and $K_{\Pi;n}(T)$ are given, in the first approximation, by Eqs. (18) and (21)-(25). Certainly, it is understood that in these expressions $D_{12}(R)$ and $U_{12}(R)$ are determined in the case of molecular ion $Ne_2^{+}$ in accordance with [61]. However, here the fact is taken into account that apart from the direct ionization channel (where the above described decay approximation is sufficient), an other, indirect ionization channel also exists. We mean the channel where $|1; R > | n >$ is considered as the initial state of the whole system and where the transition $|1; R > | n > \rightarrow | 2; R > | n-1 >$, caused by the same DRM can be realized as the first step. After that the ionization in the indirect channel can be described as in the direct one with the same value of the angular momentum, but with the new principle quantum number and collision energy, i.e. $n-1$ and $E + (\varepsilon_n - \varepsilon_{n-1})$, where $\varepsilon_{n,n-1}$ denotes the corresponding energy of RE. Therefore $K_{\Sigma;n}(T)$ and $K_{\Pi;n}(T)$ are given here by the relations

$$K_{\Sigma;n}(T) = K_{\Sigma;n}^{(dir)}(T) + K_{\Sigma;n}^{(indir)}(T) \ , \ K_{\Pi;n}(T) = K_{\Pi;n}^{(dir)}(T) + K_{\Pi;n}^{(indir)}(T)$$

(28)

where $K_{\Sigma;n}^{(dir)}(T)$, $K_{\Sigma;n}^{(indir)}(T)$, $K_{\Pi;n}^{(dir)}(T)$ and $K_{\Pi;n}^{(indir)}(T)$ are the corresponding rate coefficients for direct and indirect channels. The results of the calculations of the total rate coefficients $K_{ci;n}(T = 300K)$ for $s-$ and $d-$ series of neon Rydberg states and partial (indirect) rate coefficients $K_{ci;n}^{(indir)} = (1/3) K_{\Sigma;n}^{(indir)} + (2/3) K_{\Pi;n}^{(indir)}$, for the same $T = 300K$ are given in Fig 8. One can see that under the considered conditions the contributions of the indirect channels are relatively small. In order to compare our results with the existing experimental data, we determined here the values of the mean thermal cross sections for the associative ionization processes in $Ne^{*}(n) + Ne$ collisions in the crossed-beams, for s- and d- series of neon Rydberg states in the region $5 \leq n_{eff} \leq 20$. These results are shown in Fig. 9 together with the corresponding experimental results from [33].

It can be seen that the associative ionization cross-section for the primary excitation of the d-state at the maximum is more than two times greater than for the s-state. Moreover, the theoretical results agree with the experiment only for the s-states with $n_{eff} > 10$. So, the theory (and the experiment) need further development.

**Non-symmetric collisions.** The results of the first experiments on chemi-ionization (see, e.g. [20]) implied that non-symmetric collisions involving Rydberg atom may be characterized by considerable values of rate coefficients $k > 10^{-10} cm^3 s^{-1}$ for the thermal range of energy. So, in [62], where the process $Rb^*(nl) + K \rightarrow KRb^+ + e$ was investigated experimentally, the measured values of the rate coefficients of the process of forming hetero-nuclear clusters was comparable by its order of magnitude with typical values of coefficients for the homoatomic process. In figure 10 [63] the summary rate coefficients and branch coefficients are presented for the chemi-ionization reaction in a non-symmetric process of $Li^*(nl) + Na$ collisions $4 \le n \le 20$. Unlike the symmetric case, in the non-symmetric one $X^{(a)}(n)$ changes non-monotonically with a maximum displaced to the region of small values of $n$. The coefficients of non-symmetric processes calculated in [63] agree with their experimental values within the error of the experiment [64].

**Binary Rydberg atom – Rydberg atom Collisions.** In this case the existence of Rydberg atom + Rydberg atom-complex opens additional channels of ionization, with creation of excited states of positive and negative ions (see figure 11). The ionization processes at collisions of two Rydberg atoms have remained a scarcely explored area of the physics of atoms. There are papers in which cross-sections of the chemi-ionization process are calculated using the method of classical trajectories and the adiabatic approximation, with separation within the classical mechanics of associative and Penning ionization channels, see e.g. [65]. In that paper the parameterized cross-section of molecular-ion formation within the temperature range of $10 - 10^3 K$ may be written as

$$\sigma(n, v_{col}) = 0.703 \cdot v_{col}^{-0.65} n^{3.35} \qquad (29)$$

In [66] the rate coefficient of ionization in the system of two rubidium Rydberg atoms with $n = 11$ was calculated within the Katsuura-Smirnov model representation. The obtained values of the associative ionization rate coefficient $K_{ai}$ coincided with the experimentally measured ones

$$K_{ai} = (1.5 \pm 0.4) \cdot 10^{-8} cm^3 s^{-1}, \; T = 60K \text{ up to a factor 2.}$$

There are in the literature also data on the chemi-ionization processes of Rydberg atoms through the channel of atomic-ions creation at the binary collisions of hydrogen atoms, alcali atoms, strontium ion, obtained within the model of dipole-dipole interaction mechanism, which works well at great inter-nuclear distances in the subthermal energy range. In such a model (Katsuura-Smirnov) the dependency of the cross-section on the values of $n$ is caused by the corresponding dependency of the Rydberg atom photo-ionization cross-section [67]. In the theoretical paper [68] also within the dipole-dipole mechanism for the process

$$A^*(nl) + A^*(nl) \rightarrow (A^*(n_1 l_1) + A^+) + e \text{ a correction was taken into account due to}$$

the interaction of the initial excited states of the considered system with its nearest ones.

The oscillating character of the ionization cross-section depending on the principal quantum number was obtained, caused by the osciallations of the frequency of atom-internal transitions upon $n$ (see Fig. 12), as well as the increase of the cross-section with a decrease of temperature. The papers [66-68] may be treated as a useful step towards a full-featured theory of collisions of two Rydberg atoms.

## 6. Channels of Negative-Ion Formation in the Experiments on Chemi-Ionization

In the atmosphere of Jupiter's satellite Io the densities of $Na$ and $Cl$ atoms are of the same order of magnitude, so that at the cross-section values of $\sigma = 10^{-15} \cdot 10^{-14} cm^2$ the reactions of chemi-ionization may go on there yielding a couple of positive and a negative ion $Na + Cl_2 \rightarrow Na^+ + Cl^-$. In the general case such processes are characterized by plurality of channels

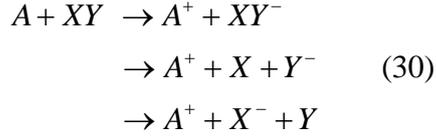

$$A + XY \rightarrow A^+ + XY^-$$
$$\rightarrow A^+ + X + Y^- \qquad (30)$$
$$\rightarrow A^+ + X^- + Y$$

Using in (30) instead of $A$ a highly excited atom $A^*(n)$ allows to explore the processes of negative-ion formation within the thermal and subthermal range of collision velocities, when the lack of kinetic energy is being compensated in the energy balance by the potential energy of the excitation. For chalogen-containing molecules with high values of the rate coefficient $K \approx 10^{-7} cm^3 s^{-1}$ the output of negative ions upon their collision with a Rydberg atom is determined by the effectiveness of the process of attachment of the RE to the chalogen-containing molecule. In the momentum approximation the cross-section for a collision of a Rydberg atom with the target molecule is determined by the cross-section of the scattering of a RE with an energy of the order of magnitude of $10^{-3} eV$ on an electro-negative molecule. For the reaction $A^*(n) + SF_6 \rightarrow A^+ + SF_6^-$, the cross-section $\sigma \sim v_{col}^{-1}$, and consequently the rate coefficient $K = <v \cdot \sigma>$ do not depend on $n$. As the Rydberg states of atoms are on the whole similar among themselves, so the lack of dependency of the rate coefficients on large $n$ is a characteristic feature of Rydberg atoms. In Fig. 13 the adapted data from [69] and [70] are given about the dependency on $n$ of the rate coefficient for the reaction $RA + SF_6 \rightarrow A^+ + SF_6^-$ within the range of the principal quantum number values $12 \le n \le 40$ for the $n^2 P$ states of a sodium atom [69] and $n^2 D$ in the case of potassium [70]. Within the limits of the experimental error $\pm 5\%$ both dependencies practically coincide.

From the relative disposition of the Coulomb term of the couple $X^+ + Y^-$ and the repulsive term of the quasi-molecule $X_2^*$ it follows that, if the inequality $1/(2n^2) < E_a$ is satisfied, where $E_a$ is the energy of the electron affinity of the atom and $1/(2n^2)$ the energy of the bond of the RE in the atom, then formation of an ion couple $A^+ + A^-$ is possible only by introduction of a third term, quasi-intersecting both initial terms. Such processes have been experimentally witnessed. In the opposite case $1/(2n^2) > E_a$ a calculation within the framework of Landau-Zenner theory yields a good agreement with the experiment [71]. Let us note that the opposite process of recombination $Na^+ + H^- \rightarrow Na^* + H$ with a recombination coefficient $\alpha = 2 \cdot 10^{-7} cm^3 s^{-1}$ leads to formation of a $Na$ atom in an excited $4^2 S$ state, that is, the upper state of electromagnetic emission at the transition $4^2 S - 3^2 P$. Other recombinative channels turn out to be less effective.

# 7. Geocosmic Aspects of Chemi-Ionization

Elementary processes in geocosmic plasma involving heavy particles traditionally attract researchers' attention, see e.g. [72-74]. Rydberg atoms and molecules are of interest for astrophysicists exploring the processes in the atmospheres of stars of late spectral classes, interstellar clouds of ionized hydrogen, where Rydberg atoms form as a result of electron-proton recombination, and other types of astrophysical plasma. Recent research [75] allow for a discussion of the influence of Rydberg molecules of atmospheric gases on the Earth's biosphere. From the research of the Solar photospheric spectra, the absorption spectra of planetary atmospheres and the chemical composition of dust it follows that alkali metals $Na$, $K$ participate in considerable amounts in the composition of cosmic objects. The composition of our Universe at its early stages of development included positive and negative atomic ions of hydrogen, deuterium, helium, lithium. The contents of this overview deals mostly with these kinds of particles.

Recent spectroscopic research of the atmospheres of cooling stars such as white dwarfs pointed out an anomaly in light emission of Rydberg atom with $n = 10$ and tabular lifetime $\tau \sim 10^{-6} s$. The lines of the corresponding infra-red transitions have not been observed [73]. Let us note that it is just these states that correspond to the maximal values of chemi-ionization rate coefficients. Accordingly to the observational data, we have that under such conditions $N_0 \geq 10^{-17} cm^{-3}$ and $N_{(*)} \geq 10^{13} cm^{-3}$, where $N_0$ and $N_{(*)}$ are the densities of the ground state and Rydberg atoms. It is not difficult to estimate that at $K_{ci:n} \approx 10^{-9} cm^3 s^{-1}$ the probability of a Rydberg atom "being extinguished" through the chemi-ionization channel is comparable to the probability of its radiative decay. The other probability, which cannot be discarded, is the manifestation of the effect of blockage of transitions akin to Förster resonance.

The interest in the model of dipole resonance mechanism in application to chemi-ionization over the last decade received a new momentum in the astrophysical problems. Namely, in [76] it was shown that the chemi-ionization processes (1) and (2) with $A = H(1s)$ in the weakly ionized hydrogen plasmas ( the ionization degree less than 0.001, $T \approx 5000K$ ) dominate over other relevant ionization processes for $4 \leq n \leq 8$ or, at least, comparable with them. Then, in [77] the similar fact was established in connection with the processes (1) and (2) with $A = He(1s^2)$ in the weakly ionized helium plasmas (the ionization degree less than 0.001, $T \approx 10000K$ ). However, the just the mentioned conditions characterize such important astrophysical object as the photospheres of the Sun and some cooler stars (for example M red dwarfs) , as well as the photospheres of some DB white dwarfs [78-80]. Keeping in mind this fact in [74] the influence of chemi-ionization processes (1) and (2) with $A = He(1s^2)$ in the reference to the other ionization processes was examined in the photospheres of the DB white dwarfs with $12000K \leq T_{eff} \leq 20000K$ . The behavior of $K_{ci;n}(T)$ for the chemi-ionization processes (1) and (2) with $A = He(1s^2)$ is illustrated in Fig. 15. On the basis of the data from [79] it was established that in the parts of the considered photosphere (where $8000 \leq T \leq 12000K$ ) these chemi-ionization processes dominate over the concurrence electron-Rydberg atom impact ionization processes.

Some later, in [81], the influence of chemi-ionization processes (1) and (2) with $A = H(1s)$ in the reference to the other concurrent ionization processes was examined in the solar photosphere. For that purpose the data from [78] were used. Together with the chemi-ionization processes with $n \geq 4$ , the similar processes with $n = 2$ , $n = 3$ were taken in to account. In this case it was also established that in the significant part of Sun photosphere the considered chemi-ionization processes dominate over the electron-Rydberg atom impact ionization processes. The values of the rate coefficients of the considered chemi-ionization processes for $2 \leq n \leq 8$ and $4000K \leq T \leq 10000K$ are given in the mentioned paper.

Apart of that, in [82,83] , by means of corresponding PHOENIX code [80], it was directly examined the influence of the chemi-ionization processes (1) and (2) with $A = H(1s)$ to the exited hydrogen atom populations, the electron density and the shape of the hydrogen spectral lines in the photosphere of M red dwarf with $T_{eff} = 3800K$. It was established that considered processes with $n \leq 8$ strongly influence to the mentioned population (for any $n > 2$ the populations changes up to $\pm 50\%$), as well as to the electron density greater than the factor $2$. As the consequence of these facts, it was established the significant change of the shape of the hydrogen atomic spectral lines (on the examples of $H_\alpha, H_\delta, H_\varepsilon$ and $Pa_\varepsilon$). In it was illustrated here by Fig. 15. These results suggest the necessity to study the chemi-ionization processes in stellar atmospheres from the spectroscopic aspect.

# 8. Influence of Non-elastic Atom – Rydberg atom Collisions on the Kinetics of Low-Temperature Laboratory Plasma

As it was shown above, large values of the rate coefficient k are characteristic of chemi-ionization processes involving Rydberg atoms in the general case. It seems natural idea to include them into the system of kinetic equations describing the state of low-temperature plasma. Lately, due to the appearance of the necessary theoretical means, for instance the apparatus of the theory of modified diffusion approximation (MDA), quantitative calculations of that kind became possible.

A nice example for this in the astrophysical applications of low-temperature plasma physics are the programs allowing modeling of stellar atmospheres of different classes at a good level and taking into account the inventory of physico-chemical reactions. For instance, the software packages Phoenix and Atlas.

The role of chemi-ionization processes in plasma kinetics is determined in the first place by the parameters of the plasma's electronic component. Thus, in a plasma with ionization non-equilibrium at an electron density $N_e \ll 10^{13} cm^{-3}$ and a fast increase of field strength the contribution of heavy particles' collisions to the ionization does not exceed 20% (inert gases). Under those conditions the plasma medium is optically translucent, the radiation leaves the plasma volume without absorption, and collisional-radiative model of plasma can omit chemi-ionization processes. At the stage of the active phase of a developed discharge (plasma of a low-voltage arc of a thermo-emissional energy converter) besides direct and cascade ionization of atoms by an electron impact, chemi-ionization processes are being taken into account $N_e = 10^{13} - 10^{16} cm^{-3}$, $T_e \geq 0.4 eV$. It is supposed that an active role in them is played by the excited atoms belonging to the block of states which are away from the ionization continuum by the magnitude of the dissociation energy of the molecular ion. Those levels practically coincide with the lower limit of an excited atom's Rydberg states. The contribution of associative-ionization processes is especially significant when a fast relaxation is possible of the ro-vibration levels of the molecular ion being formed. In the contrary case the associative ionization and its reverse process of dissociative recombination may go through the same channel (in the opposite directions) and their influence on the kinetics of ionization in the plasma diminishes.

The processes of chemi-ionization influence the form of the distribution function of the plasma electrons over energy, increase the role of cascade ionization processes, lead to forming of an anomalously large parietal layer. All this falls under the general name of the "opto-galvanic" (opto-electircal) effect, leading to a change in the external parameters of a discharge due to a change in ionizational, electric and vibrational properties of the plasma. The case of light-induced current in the rarified gas should be mentioned separately, when due to Doppler effect the excited particles being formed have a directed velocity.

Here and above the effects caused by the interaction of laser radiation with a gaseous medium were not discussed, such as spin orientation, arrangement, polarization effects, etc. In the low-temperature non-equilibrium plasma these effects scarcely affect the plasma parameters. At the same time, their physical interpretation does not in contradict the principal results obtained by means the model of the dipole resonance mechanism.

In the afterglow plasma the chemi-ionization processes become the main ionization channel, which involve metastable atoms and molecules, as well as atoms in their resonant states under conditions of capturing the resonance radiation.

**Photoplasma.** A new physical phenomenon of "photoplasma", or "photo-resonant plasma" entered the register of the contemporary plasma physics after the experiment [84], in which quasi-stationary plasma was obtained with $N_e = 10^{12} cm^{-3}$, when exposing cesium vapor to the light from gas-discharge plasma in cesium vapor. Further research on alcali-metal vapor with a density of normal atoms of $10^{15} \leq N_0 \leq 10^{17} cm^{-3}$, exposed to the light of the resonant lines with radiation power density of $10^5 \leq I \leq 10^6 Wcm^{-2}$, has shown that one of the channels of formation of slow primary electrons is the process of associative ionization $A^* + A^* \rightarrow A_2^* + e$. In figure 16 the scheme is presented of development of such photoplasma after the initial act of absorption of light on a resonant transition in the atom [85] . The lower block of processes corresponds to heating of the primary slow electrons through super-elastic collisions of the second kind with excited atoms and the subsequent processes of excitation and ionization with fast electrons.

## Conclusions

The existing experimental basis of research on chemi-ionization involving Rydberg atoms allows us to compare the results of experiment and theory at the quantitative level within the range of values $5 \leq n_{eff} \leq 25$. This primarily applies to the case of hydrogen-like alkali atoms, hydrogen itself , and some of rare-gas atoms ( $He$, $Ne$ ).

In the calculation of the chemi-ionization processes rate coefficients the dipole resonant mechanism is widely used. It is established that such a mechanism in general describes the chemi-ionization processes correctly, if the properties of the collision kinetics in various kinds of experiments are taken into account adequately.

The effect of temporal evolution of Rydberg electron in the energy space in a sigle atom-atom collision leads to multiplication of the number of possible chemi-ionization channels and may be, in principle, used in the schemes of "governing" the elementary processes of atom-atom collisions.

The further development of research in this direction must be connected with improving the precision of description of ionization effective cross-section's dependency on the value of the orbital quantum number $l$ of the excited atom and the further investigations of the non-symmetric processes (theory and experiment). An extension is particularly important of the investigation of the chemi-ionization processes in different stellar atmospheres as the factors which influence their spectral characteristics. The same refers to the role of the chemi-ionization processes in atmospheres of some planets (Io, for example)

Also, one of the very important directions for further investigations of the chemi-ionization processes is development of completely quantum-mechanical methods of their description which could be applied in the cases of extremely low temperatures.

## Acknowledgments

The authors are thankful to the Faculty of Physics at St. Peterburg State University, as well as to the Ministry of Education and Science of the Republic of Serbia for support within the Projects 176002 and III44002.

The authors would like to express their gratitude to professor N. N. Bezuglov (Faculty of Physics at St Peterburg State University), in cooperation with whom a considerable part of the original research was conducted the results of which are cited in the paper, and to professor Y. N. Gnedin (Main Astrophysical Observatory, Pulkovo, RAS, Russia) for discussion of geocosmical aspects of the processes involving Rydberg atoms.

## References

1. M. A. Morrison, E. G. Layton, and G. A. Parker (2000). Phys. Rev. Lett. **84**, 1415–1418.

2. I. I Ryabtsev, D. B Tretyakov and I. I Beterov (2003). J. Phys. B: At. Mol. Opt. Phys. **36** 297–306.

3. R. Heidemann, U. Raitzsch, V. Bendkowsky, B. Butscher, R. Löw and T. Pfau, (2008). Physical Review Letters **100**, 033601.

4. U. Raitzsch, V. Bendkowsky, R. Heidemann, B. Butscher, R. Löw and T. Pfau, (2007). Physical Review Letters **100**, 013002

5. N. Henkel, R. Nath and T.Pohl (2010). Physical Review Letters 104,id. 195302.

6. S. Badiei and L. Holmlid (2002). Chemical Physics 282: 137–146.

7. V.I. Yarygin, V.N. Sidel'nikov, I.I. Kasikov, V.S. Mironov, and S.M. Tulin (2003). JETP Letters 77, pp. 280-284.

8. L. Holmlid (2007) Journal of Physics: Condensed Matter 19: 276206.

9. M.I. Ojovan. J. Clust. Sci., http://www.springerlink.com/content/x6485w7737j11231/ (2011), doi:10.1007/s10876.011.0410.6

10.  E.A. Manykin, B.B. Zelener, B.V. Zelener. JETP Letters (2010), **92**,  p. 696-712

11.  G. V. Golubkov,  A. Z. Devdariani and  M. G. Golubkov (2002). Journal of Experimental and Theoretical Physics, **95**, pp. 987-997.

12. I. L. Beigman and  V. S. Lebedev (1995). Physics Reports **250**, pp. 95-328.

13.  J.E.Bayfield, in The Physics of Electronic and Atomic Collisions: Invited Lectures and Progress Reports given at the 9[th] International Conference on the Electronic and Atomic Collisions, edited by J. Risley and R. Geballe (University of Washington, Seattle, 1975 ), p. 726.

14. R.F. Stebbings, in Electronic and Atomic Collisions: Invited Papers, edited by G. Watel (North-Holland, Amsterdam, 1978), p. 549.

15.  I.C.Percival, in Electronic and Atomic Collisions: Invited Papers, edited by G. Watel (North-Holland, Amsterdam, 1978), p. 569.

16. I.Gersten (1976). Phys. Rev. A **14**, 1354.

17. J. A. Hornbeck and J. P. Molnar (1951). Physical Review  **84**, pp. 621-625.

18.  A. N. Klyucharev and N. S. Ryazanov (1971). Opt. Spektrosk. **31**,347.

19. Y. P. Korchevoi (1978). Zh. Eksp. Teor. Fiz. **75**, 1231.

20.  A. N. Klyucharev and  A. V. Lazarenko (1980).  Optics and Spectroscopy **48** pp.229-230.

21.  J. Boulmer, R. Bonanno and J. Weiner (1983). Journal of Physics B: Atomic **16,** pp. 3015-3024.

22.  N. N. Bezuglov, V. M. Borodin, A. K. Kazanskiy, A. N. Klyucharev, A. A. Matveev, and K. V. Orlovskiy (2001). Opt. Spectrosc. **91**, 19.

23. N. N. Bezuglov, V. M. Borodin, A. N. Klyucharev et al. (2002). Russian Journal of Phys. Chem. **76**, p. 27.

24. N. N. Bezuglov, V. M. Borodin, V. Grushevskii, A. N. Klyucharev, K. Michulis, F. Fuzo, and M. Allegrini (2003). Opt. Spectrosc. **95**, 515.

25.  M. T. Djerad, H. Harima and Cheret, M.(1985). Journal of Physics B: Atomic, Molecular, and Optical Physics **18**, pp. L815-L819.

26.  M. T. Djerad, M. Cheret and F. Gounand (1987). Journal of Physics B: Atomic, Molecular, and Optical Physics, **20**,pp. 3789-3799.

27. B C Johnson, M.-X.Wang and J. Weiner (1988). Journal of Physics B **21**, 2599.

28.  C. B. Collins, B. W. Johnson and M. J. Shaw (1972). Journal of Chemical Physics, **57**, p.5310-5316.

29.  S. Kubota, C. Davies and T. A. King (1975) Physical Review A **11**, pp.1200-1204.

30.  A. Hitachi,  C. Davies, T. A. King, S. Kubota and T. Doke, (1980). Physical Review A, **22**, pp.856-862.

31.  A.Pesnelle and S.Runge (1984). Journal of Physics B **17**, pp. 4689-4700.

 32. S.Runge, A.Pesnelle, M.Perdrix, G. Watel and  J. S.Cohen, (1985). Physical Review A **32**, pp.1412-1423.

33.  K. Harth, H. Hotop, and M. W. Ruf, in Highly Excited States of Atoms and Molecules, Invited Papers of the Oji International Seminar, Fuji-Yoshida, 1986, edited by S. S. Kano and M. Matsuzawa (University of Electro-Communications, Chofu, Tokyo, 1986), p. 117.

34.  V. M. Smirnov and A. A. Mihajlov (1971). Opt. Spectrosc. **30**, 984 (Opt. Spectrosc., USSR **5**, 525).

35. V. Ya. Aleksandrov, D. B. Gurevich, and I. V. Podmoshenskii, (1969). Opt. Spectrosc. **26**, 18. [Opt. Spektrosk. 26, 36 (1969) - in USSR]

36. E. Fermi, Nuovo Cimento 11 (1934) 157.

37.  A. Z. Devdariani, A. N. Klycharev, A. V. Lazarenko and V. A. Sheverev (1978).  Sov. Tech. Phys. Lett. **4**(9), 40.

38.  R. K. Janev, and  A. A. Mihajlov (1979). Physical Review A (General Physics) **20**, pp.1890-1904.

39.  R. K. Janev and A. A. Mihajlov (1980). Physical Review A (General Physics)  **21**,  pp.819-826

40. Lj. M. Ignjatovic and A. A.Mihajlov (2005). Physical Review A **72**, 022715.

41. E. L. Duman and I. P. Shmatov (1980). Journal of Experimental and Theoretical Physics **51**, p.1061.

42. A. A. Mihajlov, and R. K. Janev (1981). Journal of Physics B: Atomic, Molecular, and Optical Physics **14**, pp. 1639-1653.

43. G.M. Zaslavsky, R.Z. Sagdeev, D.A. Usikov and A.A. Chernikov, Weak chaos and quasi-regular patterns., (Nauka, Moscow 1993) in Russian.

44. A. Z. Devdariani, A. N. Klyucharev, N. P. Penkin and Yu. N. Sebyakin (1988). Optics and Spectroscopy **64**, pp.425-426

45. N. B. Delone, B. P. Krainov and D. L. Shepelianskii (1983). Soviet Physics - Uspekhi **26**, p. 551-572.

46. P. M. Koch and K. A. H. van Leeuwen (1995). Physics Reports **255**, p. 289-403

47. N. N. Bezuglov, V. M. Borodin, A. Eckers and A. N. Klyucharev (2002). Optics and Spectroscopy **93**, pp.661-669.

48. B. V. Chirikov (1979). Physics Reports **52**, p. 263-379.

49. I. M. Beterov and I. I. Ryabtsev (1995). AIP Conf. Proc. **329**, pp. 161-164

50. M.Yu. Zakharov, N. N. Bezuglov, A. N. Klucharev et al. in press

51. N. N. Bezuglov and A. N. Klyucharev (1979). Journal of Applied Spectroscopy **30**, pp.387-389

52. A. de Jong and F. van der Valk (1979). Journal of Physics B - Atomic and Molecular Physics. **12**, p. L561-L566.

53. J. G. Kircz, R. Morgenstern and G. Nienhuis (1982). Physical Review Letters **48**, pp.610-613

54. A. N. Klucharev (1993). Phys.-Usp. **36**, 486-512.

55. N. N. Bezuglov, A. N. Klucharev and V. A. Sheverev (1984). Journal of Physics B: Atomic, Molecular, and Optical Physics **17,** pp. L449-L452.

56.  N. N. Bezuglov, A. N. Klucharev and V. A. Sheverev, (1987). Journal of Physics B - Atomic and Molecular Physics **20**, p. 2497-2513.

57.  K. Miculis, I. I. Beterov, N. N. Bezuglov, I. I. Ryabtsev, D. B. Tretyakov, A. Ekers and  A. N. Klucharev (2005). Journal of Physics B: Atomic, Molecular, and Optical Physics **38**, pp. 1811-1831.

58.  J. Weiner and  J. Boulmer (1986).  J. Phys. B: At. Mol. Phys. **19**,  599.

59.  I. I. Beterov, D. B. Tretyakov, I. I. Ryabtsev,  N. N. Bezuglov, K. Miculis,  A. Ekers and A. N. Klucharev (2005). Journal of Physics B: Atomic, Molecular, and Optical Physics **38**, pp. 4349-4361.

60.  P. O'Keeffe,  P. Bolognesi, R. Richter, A. Moise and L.Avaldi  (2009). Journal of Physics: Conference Series, **194**, pp. 022050.

61.  T.-K.Ha, P. Rupper, A. Wüest and F.Merkt (2003). Molecular Physics **101**, p.827-838.

62.  M. T. Djera, H. Harima and  M. Cheret (1985) Journal of Physics B: Atomic, Molecular, and Optical Physics **18**, pp. L815-L819.

63.  Lj M. Ignjatovic, A. A. Mihajlov and  A. N. Klyucharev (2008). Journal of Physics B: Atomic, Molecular, and Optical Physics **41**, pp. 025203.

64. B. C.  Johnson, M. –X.  Wang and J. Weiner (1988). J. Phys. B **21,** 2599–607.

65.  M. W. McGeoch, R. E. Schlier and  G. K. Chawla (1988). Physical Review Letters  **61**,  p. 2088-2091.

66.  V. M. Borodin,  B. V. Dobrolezh,  A. N. Klyucharev and  A. B. Tsyganov (1995). Optics and Spectroscopy  **78**,  pp.15-19.

67.  N. N. Bezuglov, V. M. Borodin,  A. N. Kyucharev and M. Allegrini (1995). Optics and Spectroscopy  **79**,  pp.680-684.

68. N. N. Bezuglov, V. M. Borodin, A. N. Klyucharev, M. Allegrini and  F. Fuso (1997). Optics and Spectroscopy  **83**, pp.338-344.

69.  I. M. Beterov, G. L. Vasilenko, I. I. Riabtsev, B. M. Smirnov and  N. V. Fateyev (1987). Zeitschrift für Physik D Atoms, Molecules and Clusters **7**, pp.55-63.

70. B. G. Zollars, C. W. Walter, F. Lu, C. B. Johnson, K. A. Smith and F. B. Dunning (1986). The Journal of Chemical Physics **84**, pp. 5589-5593.

71. L. Barbier, M. T. Djerad and M. Cheret, (1986). Physical Review A (General Physics) **34**, pp. 2710-2718.

72. A. N. Klyucharev, N. N. Bezuglov, A. A. Matveev, A. A. Mihajlov, Lj. M. Ignjatovic and M. S. Dimitrijevic (2007) New Astronomy Reviews **51**, pp. 547-562.

73. Yu. N. Gnedin, A. A. Mihajlov, Lj. M. Ignjatovic, N. M. Sakan, V. A. Sreckovic, M. Yu. Zakharov, N. N. Bezuglov and A. N. Klycharev (2009). New Astronomy Reviews, **53**, p. 259-265.

74. A. A. Mihajlov, Lj. M. Ignjatovic, M. S. Dimitrijevic and Z. Djuric (2003). The Astrophysical Journal Supplement Series **147**, pp. 369-377.

75. K. Harth, H. Hotop and M.W. Ruf (2004). In: Book of Abstracts of International Conference on Problems of Geocosmos, St. Petersburg, Russia, p. 4.

76. A. A.Mihajlov, M. S. Dimitrijevic and Z. Djuric (1996). Physica Scripta, **53**, pp. 159-166.

77. A. A.Mihajlov, Z.Djuric, M. S. Dimitrijevic and N. N. Ljepojevic, (1997). Physica Scripta, **56**, pp. 631-639.

78. J. Vernazza, E. Avrett, and R. Loser (1981). ApJS, **45**, 635.

79. D.Koester (1980). A&AS **39**, 401.

80. P. H. Hauschildt, F. Allard and E. Baron (1999). ApJ, **512**, 377.

81. A. A. Mihajlov, Lj. M. Ignjatovic, V. A. Sreckovic and M. S.Dimitrijevic (2011). ApJS **193**, 2.

82. A. A. Mihajlov, D. Jevremovic, P. Hauschildt, M. S. Dimitrijevic, Lj. M. Ignjatovic and F. Alard (2003). Astronomy and Astrophysics, **403**, p.787-791.

83. A. A. Mihajlov, D. Jevremovic, P. Hauschildt, M. S. Dimitrijevic, Lj. M. Ignjatovic and F. Alard (2007). Astronomy and Astrophysics **471**, pp.671-673.

84. N. D. Morgulis, Yu. P. Korchevoi and A. M. Przhonskii (1968). Soviet Physics JETP **26**, p.279.

85.  N. N. Bezuglov, A. N. Klyucharev and T. Stacewicz (1994). Optics and Spectroscopy **77**, pp.304-326.

# Figure captions

Fig. 1 Schematic representation of chemi-ionization processes in decay approximation. Chemi-ionization processes in a collision complex A* (n) + A in decay approximation.

Fig. 2 Potential curves of a symmetric atom-Rydberg atom system, A* (n) + A, as functions of the internuclear distance R.

Fig. 3 Dependency of energy of the external electron in a Rydberg atom on the duration of stochastic diffusion:

(a) the case of global dynamic chaos,

(b) regime of partial optical-transition blockage (Ferster resonance).

Fig. 4 Particle distribution functions over the relative speeds in the cases of:

(sb)   -  single atomic beam;

(c)     -  gas cell (plasma);

(cb)   - orthogonal crossed beams;

(cb1)  -  counter beams;

$v_0 = \left( kT / \mu \right)^{1/2}$, where T is the effective temperature in the reaction zone, and μ - the reduced mass of the particles [56].

Fig. 5  The rate coefficients for associative ionization in Na*(n2p)+Na collisions; orthogonal crossed beams – experiment, cell – theory ; T = 600 K [72].

Fig. 6  Associative ionization Na*(neff, l) + Na rate coefficients:

(a) experiment – dots, l = 1, T = 600 K [21] ;

(b) single atomic beam, l = 1, T =1 000 K [58];

(c) single atomic beam, l = 2, T =1 000 K [58];

(d) single atomic beam, l = 0, T =1 000 K [58]; theory [57].

Fig. 7 Rate coefficients for associative ionization in $Na^*(n)+Na$ collisions determined in orthogonal crossed beam after the separation of the contribution of "catching-up" collisions from the ion signal. Continuous line – theory [59].

Fig. 8 The total rate coefficients for the chemi-ionization in Ne*(neff , l) + Ne collisions

Fig. 9  Cross-sections for associative ionization in symmetric collisions of Rydberg atoms Ne*(neff , l) in S- and D-states with normal atoms: experiment – [33], theory .

Fig. 10  Total rate coefficients for the chemi-ionization (A I + P I ) in non-symmetric Li*(n, l) + Na collisions (l = 1), and branch coefficients X(a)(n, T) in orthogonal crossed beams  experiment [63].

Fig. 11  Illustration of quasi-intersection of the potential curves of the systems created in binary collisions of Rydberg atoms.

Fig. 12  Cross sections $\sigma(n)$ for the process of  associative-ionization in A*(n) + A collisions [68].

Fig. 13   Rate coefficients K(n) for the collision process A*(n) + SF6 -> A+ + SF6- : data from  [69].

Fig. 14   Theoretical total rate coefficients for the chemi-ionization processes He*(n) + He collisions [74].

Fig. 15   The influence of the chemi-ionization processes to the shape of hydrogen atomic spectral lines $H_\delta$ and $H_\varepsilon$: full line-  with these processes,  thin line- without  of them [73].

Fig. 16 The scheme of creation of currentless photoresonant plasma.

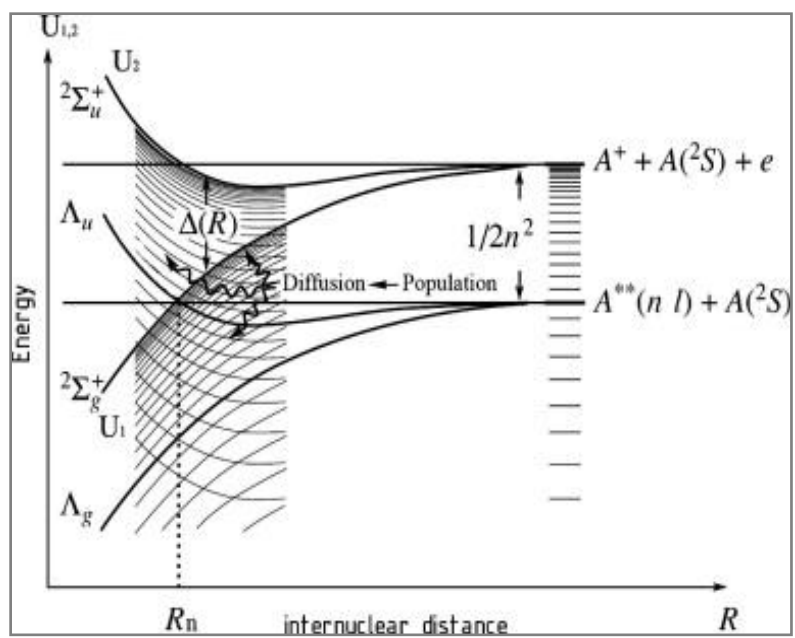

Fig1

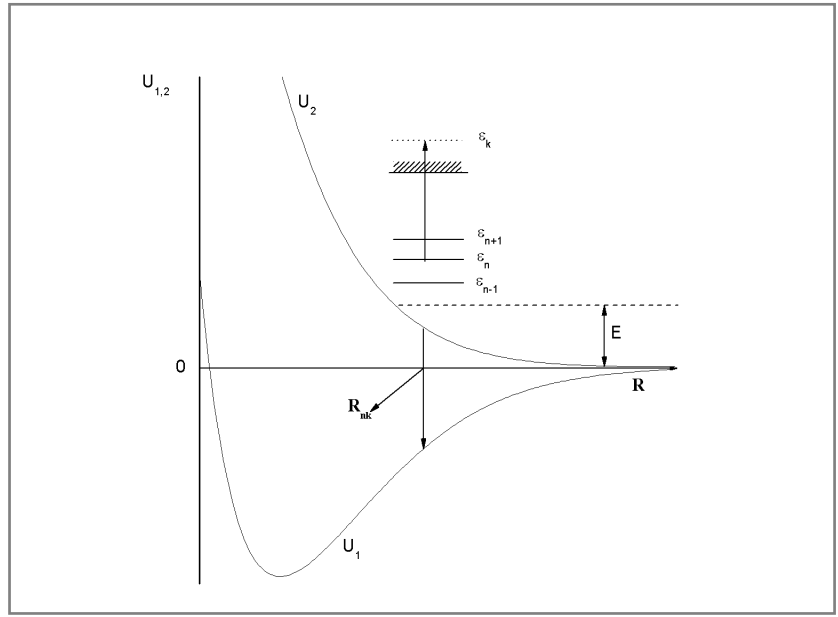

Fig2

Fig3

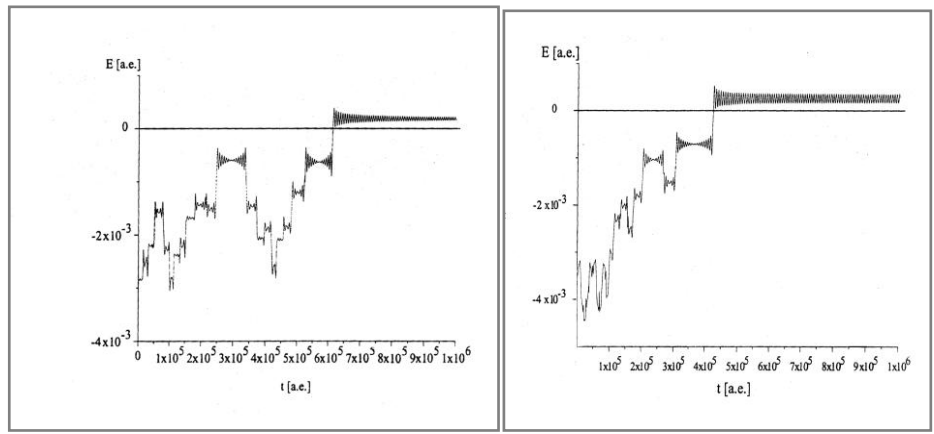

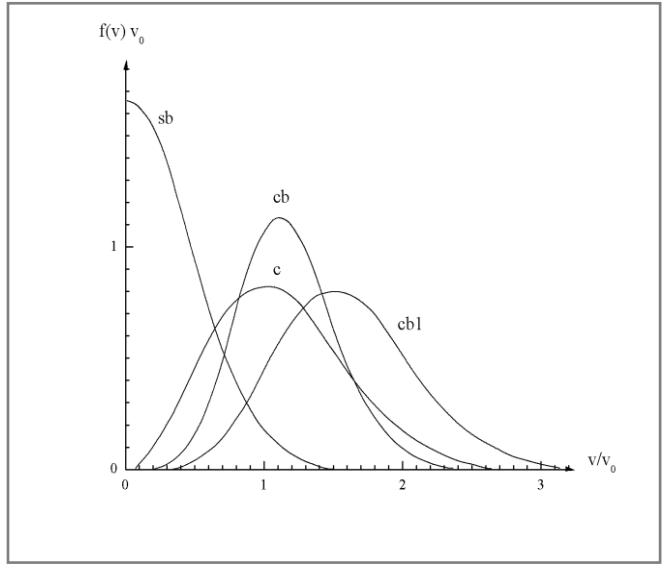

Fig4

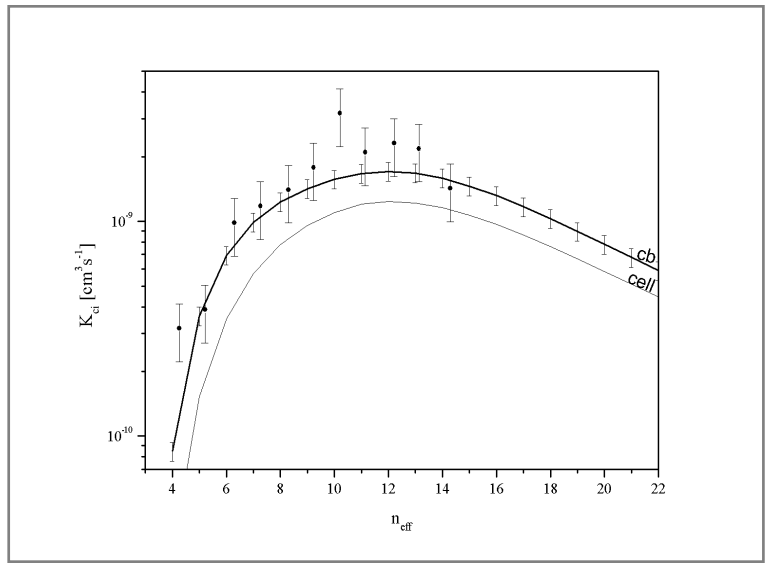

Fig5

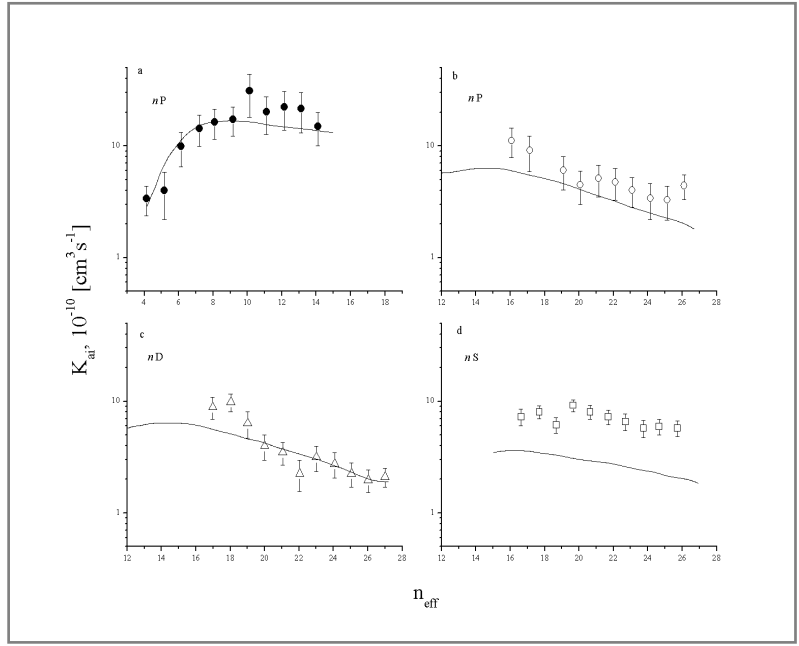

Fig6

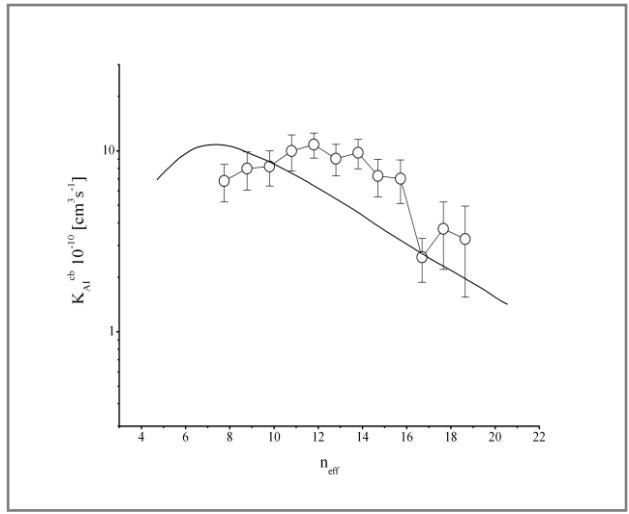

Fig7

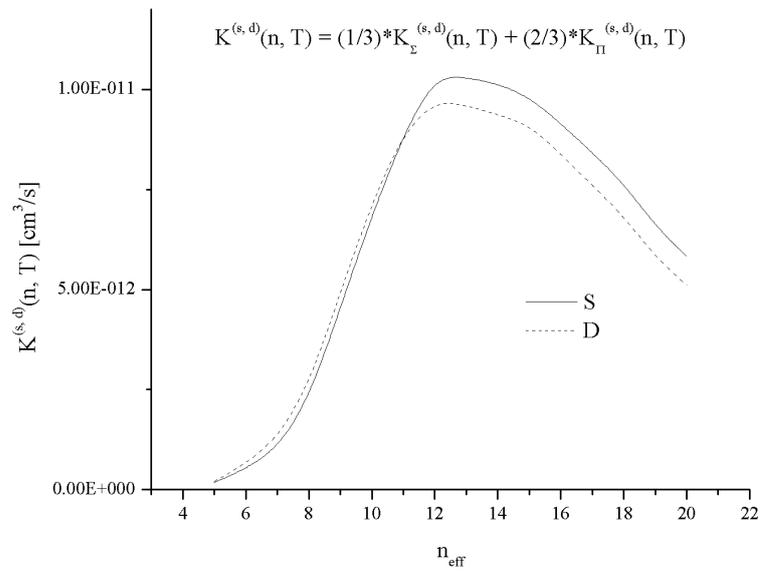

$$K^{(s,\,d)}(n,\,T) = (1/3)*K_{\Sigma}^{(s,\,d)}(n,\,T) + (2/3)*K_{\Pi}^{(s,\,d)}(n,\,T)$$

Fig8

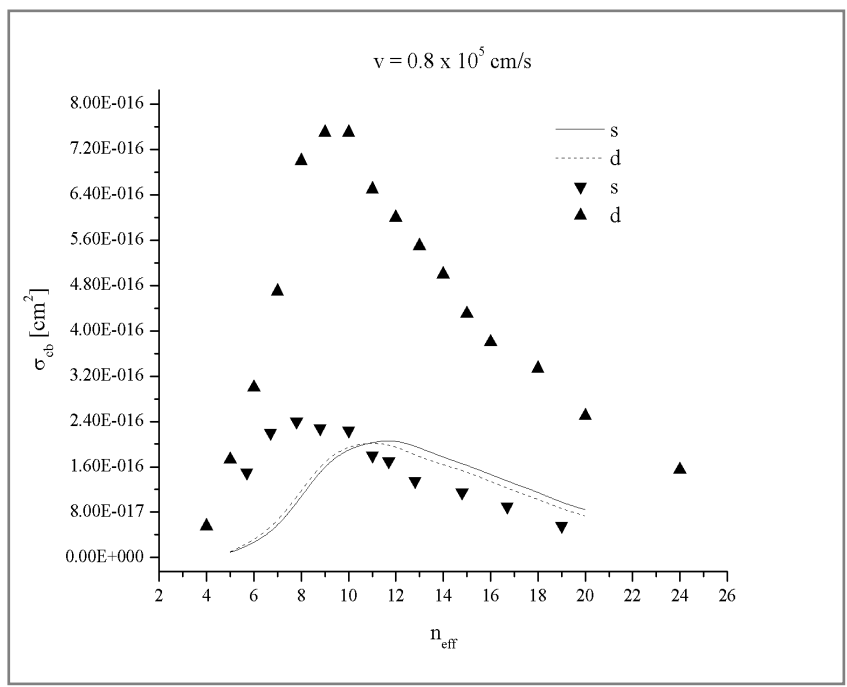

Fig9

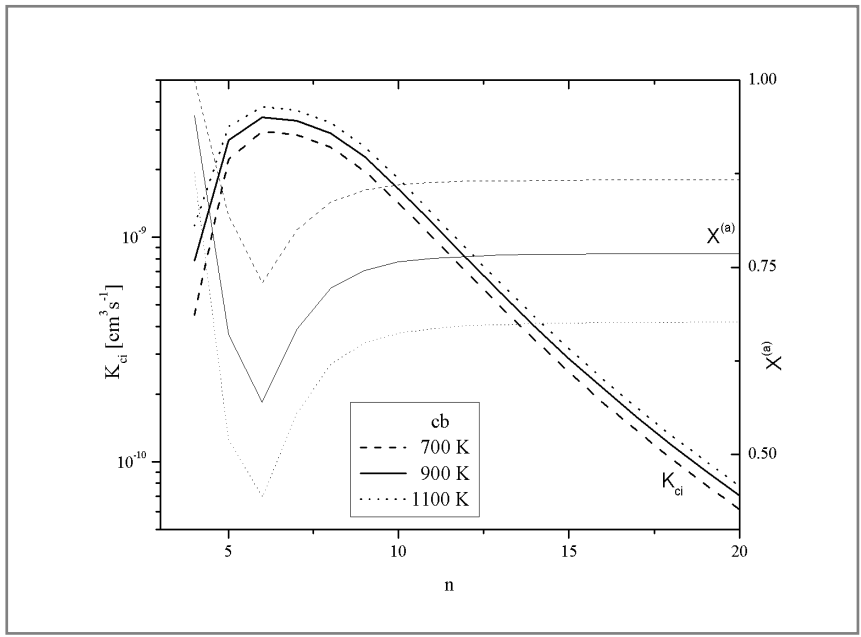

Fig10

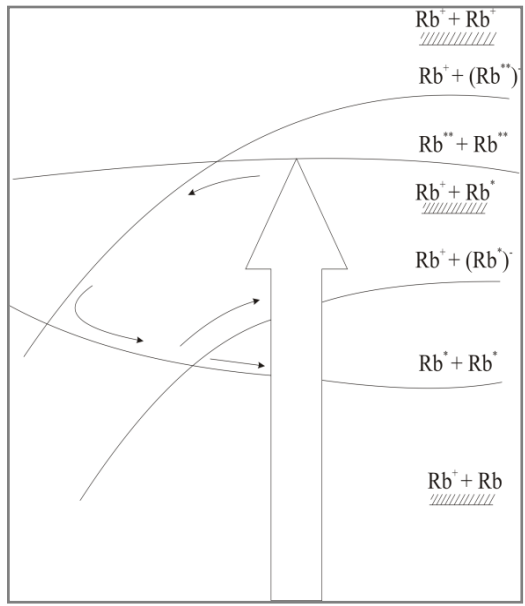

Fig11

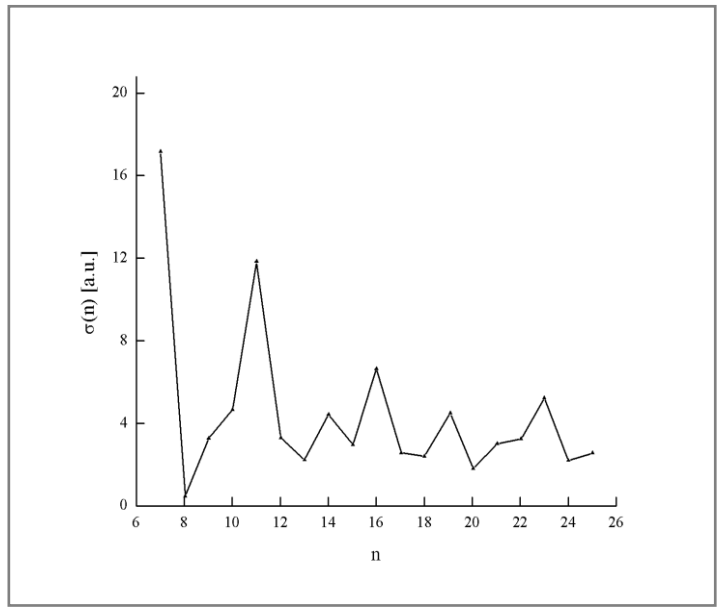

Fig12

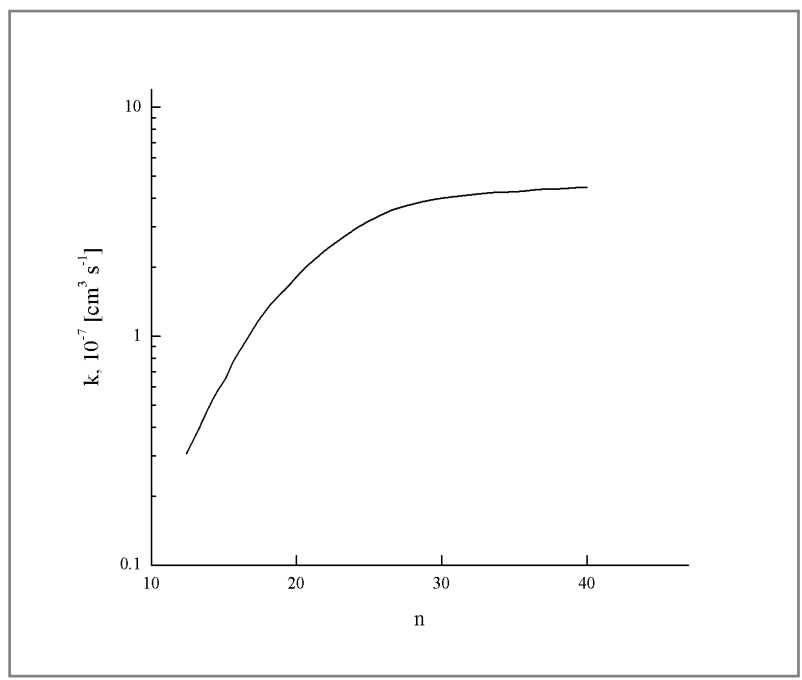

Fig13

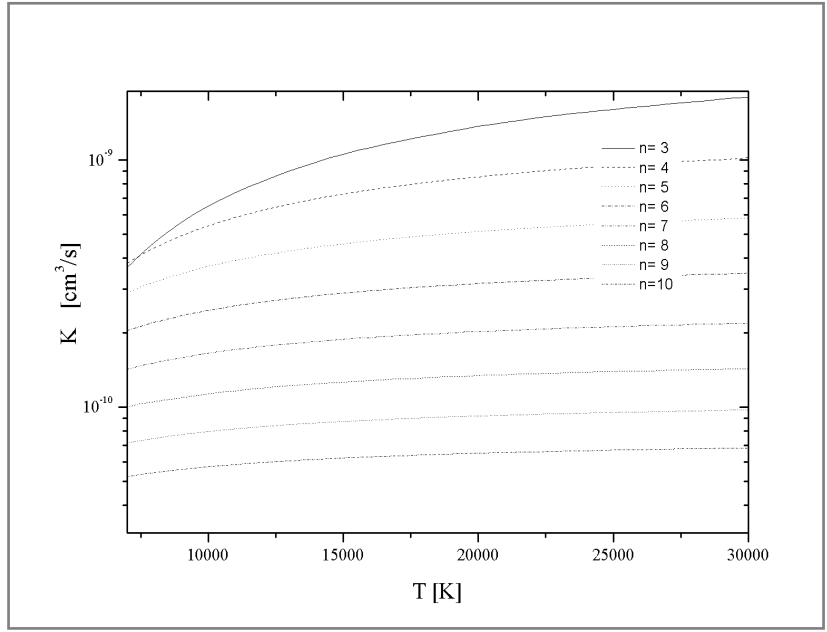

Fig14

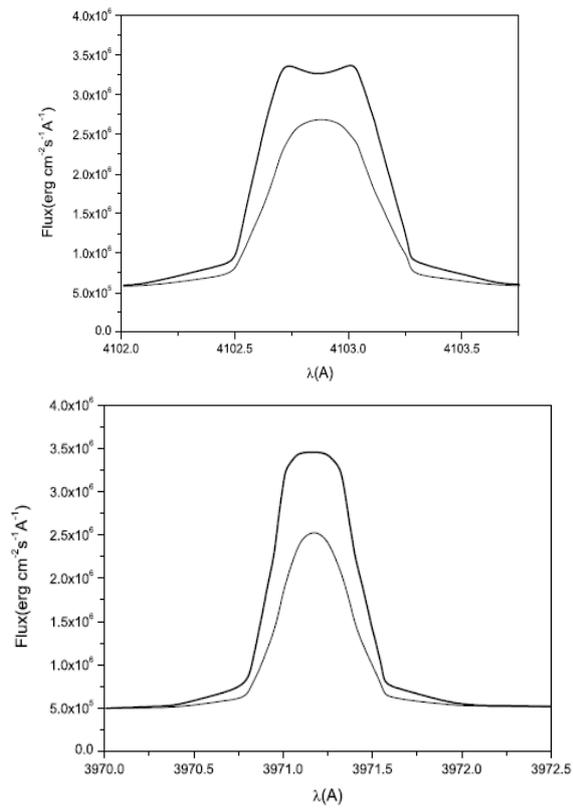

Fig15

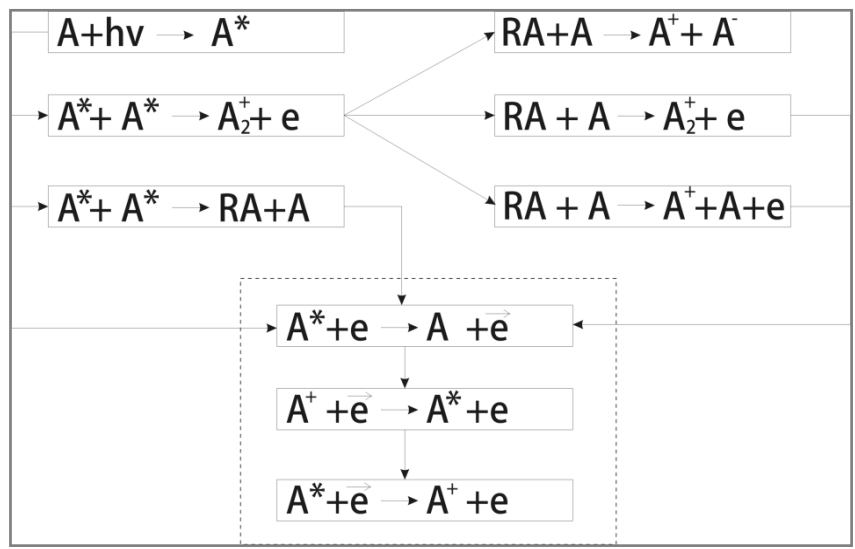

Fig16